\documentclass[prb,preprint,superscriptaddress,groupedaddress,showpacs,preprintnumbers,amsmath,amssymb]{revtex4}
\usepackage{graphicx}
\usepackage{subfigure}

\begin{document}
 
\title{Eigenmode decomposition of the near-field enhancement in localized 
surface plasmon resonances of metallic nanoparticles}

\author{Titus Sandu}
\affiliation{National Institute for Research and Development in Microtechnologies,
126A, Erou Iancu Nicolae street, 077190, Bucharest, ROMANIA}
\email{titus.sandu@imt.ro}
%\date{\today}

\begin{abstract}
I present a direct and intuitive eigenmode method that evaluates 
the near-field enhancement around the surface of metallic nanoparticles of arbitrary shape. The method is  
based on the boundary integral equation in the electrostatic limit. Besides the nanoparticle 
polarizability and the far-field response, the near-field enhancement around nanoparticles can be also conveniently expressed as 
an eigenmode sum of resonant terms. Moreover, the spatial configuration of the near-field enhancement depends explicitly on the 
eigenfunctions of both the BIE integral 
operator and of its adjoint. It is also established a direct physical meaning of the two types of eigenfunctions. 
While it is well known that the eigenfunctions of the BIE operator are electric charge modes, it is less known and used that the eigenfunctions 
of the adjoint represent the electric potential generated by the charge modes. 
For the enhanced spectroscopies the present method allows an easy identification 
of hot spots which are located in the regions with maximum charge densities and/or regions with fast 
variations of the electric potential generated by the charge modes on the surface.
This study also clarifies the similarities and the differences 
between the far-field and the near-field behavior of plasmonic systems. Finally, the analysis of concrete examples like the 
nearly touching dimer, the prolate spheroid, and the nanorod illustrate some modalities to improve the near-field enhancement.
\end{abstract}

\maketitle
\section{Introduction}

The interaction of light with conduction electrons in metallic nanoparticles 
(NPs) results in localized surface plasmon resonances (LSPRs) that have the 
ability to guide, manipulate, and enhance light fields \cite{Maier2007}. The LSPRs are 
typically confined to length scales much smaller than the diffraction limit, 
which makes them suitable for localization and enhancement of 
electromagnetic fields. These properties enable applications in sensing \cite{Mayer2011}, 
waveguiding \cite{Lal2007}, optical information processing \cite{Engheta2007}, or photovoltaics \cite{Atwater2010}. 
Particularly, the near-field enhancement is exploited in near-field 
microscopy \cite{Mock2002}, photoluminescence \cite{Tam2007}, higher harmonic generation \cite{Danckwerts2007,Kim2008}, 
and in several enhanced spectroscopies like enhanced fluorescence spectroscopy \cite{Kinkhabwala2009}, 
surface enhanced Raman spectroscopy (SERS) \cite{Nie1997,Kneipp2008,Lorenzo2009}, and surface 
enhanced Infrared spectroscopy (SEIRS) \cite{Le2008,Neubrech2008,Adato2009}. 

Numerical and theoretical methods used to predict and calculate the 
properties of LSPRs are successfully based on the integration of 
Maxwell's equations. The finite-difference time domain method \cite{Oubre2004}, 
the discrete-dipole approximation \cite{Draine1994}, and the boundary element 
method  \cite{Abajo2002,Hohenester2005} are typical computational methods for 
full electromagnetic calculations of the 
optical response in metallic NPs. These complex numerical schemes 
present, however, little intuitive help about the nature and the physics of 
the LSPRs with respect to parameters like the shape (geometry) or complex 
dielectric functions of nanoparticles. The hybridization model has 
been proposed as an alternative approach which works very well in the 
quasi-static limit \cite{Prodan2003}. This model offers an intuitive physical 
picture in terms of plasmon eigenmodes. On the other hand, in the 
quasi-static limit, the LPSRs are in fact electrostatic resonances of a 
linear response operator \cite{Fredkin2003}, which is defined on the boundary of the NP 
resulting a boundary integral equation (BIE) for an arbitrary 
geometry. This linear response operator and its adjoint, the 
Neumann-Poincare operator, are associated with the Neumann and Dirichlet 
problems in potential theory, respectively \cite{Kellogg1929,Putinar2007}. In essence, the BIE method relates the 
LSPRs to the eigenmodes of the linear response operator and the 
Neumann-Poincare operator, such that the spectral studies of the linear 
response operator provide useful information about the LSPRs. The BIE method may 
work perturbatively even beyond the quasistatic limit \cite{Mayergoyz2005,Pedersen2011}. Moreover, being 
able to calculate the polarizability of a generic dielectric particle, the 
method can be applied not only to LSPRs in metallic NPs, but also for 
polarizability calculations of biological cells in radiofrequency \cite{Sandu2010}. Like 
the hybridization model, the spectral approach to BIE offers the same 
advantages of intuitive view of plasmon eigenmodes. The method can be 
extended to clusters of NPs \cite{Davis2009} such that symmetry and selection rules that 
are used in the hybridization model \cite{Brandl2006} can be applied directly in BIE \cite{Zhang2010,Gomez2010} 
by considering the cluster eigenmodes as hybridizations of individual NP eigenmodes \cite{Davis2010}. 

Factors like composition, size, geometry, as well as the embedding media 
determine the LSPRs of metallic NP \cite{Noguez2007}. In many applications there is a 
need for precise locations of the LSPRs. In SEIRS applications, for example, 
the spectral localization of the LSPR needs to be as close as possible to 
the molecular vibration that is to be enhanced and therefore sensed. In 
addition to that, the near-field enhancement factor is a key figure of merit 
in the enhanced spectroscopies where the geometry plays an important role. 
Large near-field enhancement occurs at a sharp tip by the lightning-rod 
effect \cite{Novotny1997} or at the junctions of NP dimers \cite{Atay2004}. The geometrical 
arrangement in dimers, as opposed to single NPs, exhibit much stronger field 
enhancements; thus, as the distance between dimer NPs decreases, the 
near-field enhancement increases in the space between the NPs of the dimer 
\cite{Atay2004,Rechberger2003,Su2003,Romero2006}. 

While the BIE method permits the  calculation of near-field 
enhancement \cite{Davis2010b} a direct and intuitive way to extract the near-field enhancement factor 
is still needed. In this work I present a method that provides explicitly the near-field enhancement and 
its spatial variation in terms of eigenfunctions of the linear response operator and its adjoint. The spatial distribution 
of the field enhancement normal to the surface of the NP is proportional to the eigenfunctions of the linear operator. 
These eigenfunctions are charge modes, therefore the near-field maxima occur at the maxima of the surface charge density. 
On the other hand, the tangential component of the near-field enhancement is proportional to the derivative of the adjoint operator 
eigenfunctions, which are, in fact, the surface electric potential generated by the charge modes. 
The latter aspect has been hardly used in plasmonic applications. In addition to that, the current method directly ascertains 
the relationship between far-field and near-field spectral properties of the LSPRs \cite{Zuloaga2011}. 
The proposed method has also limitations. First, it is valid only in the quasistatic approximation, 
hence the NPs must be much smaller than the light wavelength. The second issue comes from the quantum nature of the LSPR phenomenon. 
Thus the electron spill-out and the nonlocality of the electron interaction determine a different electric field behavior at the 
surface of the NP\cite{Abajo2008, Zuloaga2009, Zuloaga2010, David2011}. However, at distances above 1 nm the classical description works well. 
It is proved by several examples that, despite these shortcomings, the present method remains a powerful tool for locating and improving 
the near-field enhancement in plasmonic systems. 

The paper has the following structure. The next section details exhaustively the method of calculating the spatial configuration of 
the near-field enhancement as depending on the eigenvalues and eigenfuntions of the BIE operators. Section III presents the numerical 
implementation and two comparative studies: the sphere versus the nearly touching dimer and the nanorod versus the prolate spheroid. 
In the last section I summarize the conclusions.

\section{Theoretical background}

For the sake of clarity I present first the main results of the spectral approach 
to the BIE method. Let us consider a NP of volume $V$ which is delimited by the surface 
$\Sigma $ and has a dielectric permittivity $\epsilon_1 \left( \omega \right)$. The 
NP is embedded in a uniform medium of permittivity $\epsilon_0 \left( \omega 
\right)$. In the quasi-static limit, i. e., the size of NP is much smaller 
than the wavelength of incident radiation, the applied field is almost 
homogeneous and the Laplace equation suffices to describe the 
behavior of the NP under the incidence of the light

\begin{equation}
\label{eq1}
\Delta \Phi ( \bf{x}) = 0,\;\;\bf{x} \in \Re ^3\backslash 
\Sigma ,
\end{equation}

\noindent
where $\Phi $ is the potential of the total electric field ${\bf {E}}_{total} $, i.e., ${\bf {E}}_{total} ({\bf{x}}) = - \nabla \Phi ( \bf{x}) = {\bf {E}} ( {\bf{x}}) + {\bf {E}}_0 ( {\bf{x}})$ and $\Re ^3$ is 
the Euclidian 3-dimensional space in which the NP of surface $\Sigma $ is embedded. The boundary conditions are: 
$ {\epsilon_0 ( \omega )\frac{\partial \Phi }{\partial \bf {n}}} |_ + = {\epsilon_1 ( \omega )\frac{\partial \Phi }{\partial \bf{n}}} |_ - $ for 
$\bf{x} \in \Sigma$; and $ - \nabla \Phi ( {\bf{x}}) \to {\bf{E}}_0 $ for $| \bf{x} | \to \infty$, 
where $\bf{n}$ is the outer normal to the surface $\Sigma $ and ${\bf{E}}_0 $ is the incident (applied) field. 
The solution of (\ref{eq1}) can be expressed as a superposition of the applied electric potential $-{\bf{x}}\cdot {\bf{E}}_0 $ 
and a single-layer potential generated by the surface charge distribution $u({\bf{x }})$, 

\begin{equation}
\label{eq3}
\Phi ( {\bf{x }}) = - {\bf{x}} \cdot {\bf{E}}_0 + \frac{1}{4\pi }\int\limits_{{\bf y} \in \Sigma } 
{\frac{u( {\bf y} )}{| {{\bf x} - {\bf y}} |}d\Sigma_{y} } .
\end{equation}

\noindent
The single layer-potential utilized in (\ref{eq3}) can define on $\Sigma $ a symmetric 
operator $\hat {S}$ that acts on the Hilbert space $L^2(\Sigma)$ of square-integrable functions on $\Sigma $ as

\begin{equation}
\label{eq4}
\hat {S}[u] = \frac{1}{4\pi }\int\limits_{
 {\bf y} \in \Sigma, {\bf x} \in \Sigma } {\frac{u\left( {\bf y} \right)}{\left| {{\bf x} - {\bf y}} \right|}d\Sigma_{y} } .
\end{equation}

\noindent
In the Hilbert space $L^2( \Sigma)$, the scalar product of two 
functions $\tilde {u}_1 ( {\bf x} )$ and $\tilde {u}_2 ( {\bf x} 
)$ is defined as

\begin{equation}
\label{eq5}
\langle {\tilde {u}_1 } | \tilde {u}_2 \rangle = 
\int\limits_{ {\bf x} \in \Sigma } {\tilde {u}_1^\ast ( {\bf x} )\tilde {u}_2 ( {\bf x})d\Sigma_{x} } .
\end{equation}

\noindent
The derivative of the single-layer potential presents discontinuities
across the boundary $\Sigma $. This can be used to rewrite (\ref{eq1}) with the help of the operator 
also defined on $L^2( \Sigma )$ \cite{Kellogg1929,Sandu2010,Putinar2007}

\begin{equation}
\label{eq6}
\hat {M}[u] = \frac{1}{4\pi }\int\limits_{
 {\bf y} \in \Sigma, {\bf x} \in \Sigma} 
{\frac{u( {\bf y} ){\bf n}_x \cdot ( {{\bf x} - {\bf y}})}{| {{\bf x} 
- {\bf y}} |^3}d\Sigma_{y} } .
\end{equation}

Then, the equation fulfilled by the charge distribution $u( {\bf x} )$ 
in Eq. (\ref{eq3}) has the following operator form

\begin{equation}
\label{eq7}
\frac{1}{2\lambda }u( {\bf x} ) - \hat {M}[ u ] = {\bf n} \cdot {\bf E}_0, 
\end{equation}

\noindent
with $\lambda = \frac{\epsilon_1 - \epsilon_0 }{\epsilon_1 + \epsilon_0 }$. The function $u( {\bf x})$, which defines 
through (\ref{eq3}) the solution of (\ref{eq1}), can be found by the knowledge of the 
eigenvalues $\chi _k $ and eigenfunctions of $\hat {M}$  and its of adjoint operator expressed as:

\begin{equation}
\label{eq9}
\hat {M}^ {\dagger} [ v ] = \frac{1}{4\pi }\int\limits_{
 {\bf y} \in \Sigma,  {\bf x} \in \Sigma} {\frac{v({\bf y} ){\bf n}_y \cdot ({{\bf x} - {\bf y}})}{| {{\bf x} 
- {\bf y}}|^3}d\Sigma_{y} } .
\end{equation}

\noindent
This is the Neumann-Poincare operator \cite{Putinar2007} and has the physical significance of an electric potential 
generated by a dipole distribution on $\Sigma$. The operators $\hat {M}$ and $\hat {M}^ {\dagger}$ have several 
general properties. Their eigenvalues 
are equal and restricted to [-1/2 ,1/2], while their eigenfunctions 
are bi-orthogonal, i.e. if $\hat {M}[{u_i}] 
= \chi _i u_i $ and $\hat {M}^{\dagger} [ {v_j }] = \chi _j v_j $, then 
$\langle {v_j } | {u_i } \rangle = \delta _{ij} $\cite{Fredkin2003,Mayergoyz2005,Sandu2010,Ouyang1989,Sandu2011}. However, the 
eigenfunctions $u_i $ and $v_i $ are coupled through the Plemelj's 
symmetrization principle \cite{Putinar2007}

\begin{equation}
\label{eq10}
\hat {M}^ {\dagger} \,\hat {S} = \hat {S}\,\hat {M}.
\end{equation}

\noindent
One can notice that the operator $\hat {M}$ can be made symmetric with 
respect to the metric defined by the symmetric and non-negative operator 
$\hat {S}$, i. e., for any $\tilde {u}_1 ,\tilde {u}_2 \in L^2\left( \Sigma 
\right)$: $\langle {\tilde {u}_1 } | {\tilde {u}_2 } \rangle _S 
= \langle {\tilde {u}_1 } |\hat {S} {[ {\tilde {u}_2 } 
]} \rangle $.
Using (\ref{eq10}) and the norm defined by $\hat {S}$ one can relate the eigenfunctions $u_i $ and $v_i $ 
by

\begin{equation}
\label{eq12}
v_i = \hat {S}\left[ {u_i } \right].
\end{equation}

\noindent
From physical point of view, Eq. (\ref{eq12}) denotes that $v_i $ is the electric potential generated on 
surface $\Sigma $ by the charge distribution $u_i $ and, to the author's knowledge, Eq. (\ref{eq12}) has not been used 
in plasmonic applications. As it will be shown later, Eq. (\ref{eq12}) is instrumental for the calculation of 
the near-field enhancement in a coordinate system directly related to the geometry of the NP.

The explicit solution of (\ref{eq7}) can be expressed in terms of eigenvalues 
and eigenfunctions of $\hat {M}$ and $\hat {M}^{\dagger} $ as \cite{Sandu2010,Sandu2011} 

\begin{equation}
\label{eq17}
u = \sum\limits_k {\frac{n_k }{\frac{1} {2\lambda } - \chi _k }} 
u_k .
\end{equation}
In (\ref{eq17}) ${n}_{k} = \langle {v}_{k} \vert {\bf{n}} \cdot 
{\bf{N}}\rangle $, where $\bf{N}$ the unit vector of the applied field given by ${\bf{E}}_0 = {E}_{0} {\bf{N}}$. The term ${n}_{k}$ is
the contribution of the k$^{th}$ eigenmode to the solution of (\ref{eq7}) and, as shown below, it represents the 
weight coefficient of the k$^{th}$ eigenmode to the evanescent near-field. Also, the charge density $u$ determines an 
electric potential $v$ on $\Sigma$  via Eq. (\ref{eq12}). In addition, (\ref{eq17}) has one part that depends on the geometry through the 
eigenfunctions and the second part that depends on both the geometry (through the eigenvalues) and the dielectric 
properties. 

The charge density $u$ determines the volume-normalized polarizability of the NP 
as the volume-normalized dipole moment generated by $u$ along the applied 
field direction \cite{Sandu2010,Sandu2011}  

\begin{equation}
\label{eq18}
\alpha = \sum\limits_k {\frac{w_k }{{\frac{1} {2\lambda }} - \chi 
_k }}, 
\end{equation}

\noindent
where $w_{k} = 
{n}_{k} \langle {\bf{r}} \cdot {\bf N}\vert {u}_{k} \rangle /V$ is the weight of the k$^{th}$ eigenmode to the NP polarizability 
and $\bf r$ is the position vector that determines $\Sigma$. The parameter ${w}_{k} < 1$ is scale-invariant and solely determined by 
the geometry of the NP. 

One may obtain explicit expressions for 
$\alpha $ if a Drude form $\epsilon = \varepsilon _{m} - {\omega _{p}^2 }/(\omega(\omega + i\gamma)) 
$ is used for the complex permittivity of metals. Here, 
$\varepsilon _{m} $ incorporates the interband transitions (with little 
variations in VIS-IR) and the term $\varepsilon _\infty 
$. The parameter $\omega _p $ is the plasma resonance frequency of free 
electrons and $\gamma$ is the Drude relaxation term. Dielectrics are in contrast 
described by a real and constant dielectric 
function $\epsilon  = \varepsilon _{d}$. Including these explicit expressions 
for the dielectric permittivities, the NP polarizability is \cite{Sandu2011}

\begin{equation}
\label{eq19}
\alpha_{plas} (\omega) = 
\sum_k \frac{w_k (\varepsilon_m - \varepsilon_{d})}{\varepsilon_{eff\_k}}-\frac{w_k}{1/2-\chi_k} \frac{\varepsilon_{d}}{\varepsilon_{eff\_k}}
\frac{\tilde\omega_{pk}^2}{\omega(\omega + i\gamma) - \tilde\omega_{pk}^2},
\end{equation}

\noindent
where $\tilde {\omega }_{pk}^2 = {( {1 / 2 - \chi _k })\omega _p^2 }/{\varepsilon _{eff\_k} }$ is the square of a frequency 
associated with the resonance of the $k^{th}$ eigenmode and $\varepsilon _{eff\_k} = ( 
{1 / 2 + \chi _k })\varepsilon _d + ( {1 / 2 - \chi _k })\varepsilon _m$ is an effective dielectric parameter. In visible and 
infrared Eq. (\ref{eq19}) has a slow-varying part and a sum of fast-varying Drude-Lorentz  
terms $ - {w_k }/{( 1 / 2 - \chi _k)} \times {\varepsilon _d }/ {\varepsilon _{eff\_k} } \times {\tilde {\omega }_{pk}^2 /} 
{( {\omega ( {\omega + i\gamma }) - \tilde {\omega }_{pk}^2 })}
 $. 

The far-field behavior of the interaction of electromagnetic fields with 
metallic NPs is determined by the induced dipole that is proportional to the 
normalized polarizability $\alpha $. The 
imaginary part of the polarizability is directly related to the absorption/extinction of light which is the far-field effect of the LSPRs. Thus the 
cross-section of the extinction is \cite{Maier2007}

\begin{equation}
\label{eq20}
C_{ext} = \frac{2\pi }{\lambda }Im\left( {\alpha_{plas} V} \right),
\end{equation}

\noindent
where $\lambda $ is the wavelength of the incident radiation. 
Now it becomes apparent that $w_k $ signifies the weight of the 
the k$^{th}$ eigenmode to the far-field of the LSPRs. The eigenmodes which have $w_k \ne 0$ 
couple with light and therefore are bright eigenmodes, whilst those which have $w_k = 0$ are 
dark eigenmodes. The eigenmode with the largest eigenvalue $1/2$ is dark because 
always $\langle {v}_{1} \vert {\bf{n}} \cdot {\bf{N}}\rangle = 0$. Physically, it represents a 
monopole charge distribution. 
The strength of each bright eigenmode is in fact proportional to the geometric factor $w_k / (1/2 - \chi _k )$, such that some eigenvalues $\chi _k $ close to $1/2$ might show strong plasmon resonance response even though ${w}_{k}$ may have low values \cite{Sandu2011}. Moreover, 
as it will be seen below, if one neglects $\gamma$, then the resonance frequency $\tilde {\omega }_{pk} $ is just the 
LSPR frequency of the k$^{th}$ eigenmode. Therefore, larger $\chi _k$'s mean longer plasmon wavelengths and, as an 
eigenvalue $\chi _k $ approaches $1/2$, the plasmon 
resonance frequency moves in the mid-infrared \cite{Sandu2011,Romero2006}.

In principle, the near-field around NP can be evaluated 
from Eq. (\ref{eq17}) by calculating first the electric potential and then the 
electric field. Below I will present compact and intuitive relations for the near-field at the surface $\Sigma$ of the NP. 
These relations allow a decomposition of the near-field at $\Sigma$ in normal and tangent components and a direct 
calculation of the near-field enhancement. For this purpose I will utilize a coordinate system directly related to the 
parameterization of surface $\Sigma$. Let us suppose that $\Sigma $ is locally 
parametrized by $x = X( {\xi^1,\xi^2})$, $y = Y({\xi^1,\xi^2 })$, and $z = Y( {\xi^1,\xi^2})$, where 
$\xi^1 ,\;\xi^2 $ are the independent parameters defining $\Sigma $. 
If the functions X, Y, and Z are sufficiently smooth, the vectors tangent to $\Sigma $ are defined by \cite{Moon1988} 

\begin{equation}
\label{eq21}
{\bf r}_{\xi^{1,2} } = {\frac{\partial {\bf r}}{\partial \xi^{1,2} }},
\end{equation}

\noindent
whose norms $h_{\xi^{1,2} } = | {{\bf r}_{\xi^{1,2} }} |$ are the Lam\'{e} coefficients. The unit vectors  
${\bf t}_{\xi^{1,2} } = {{\bf r}_{\xi^{1,2} }} / h_{\xi^{1,2} }$
determine the normal on $\Sigma $ as the cross-product

\begin{equation}
\label{eq22}
{\bf n} = {\bf t}_{\xi^1 } \times {\bf t}_{\xi^2 } /|{\bf t}_{\xi^1 } \times {\bf t}_{\xi^2 }|.
\end{equation}

\noindent
In Eqs. (\ref{eq21}) and (\ref{eq22}) the position vector ${\bf r}$ designates a point on $\Sigma $ and 
therefore depends on $\xi^1 $ and $\xi^2 $. The following nonlinear coordinate 
transformation $( {x,y,z}) \leftrightarrow ( {\xi^1 ,\xi^2 ,\xi^3 } )$ allows the decomposition of the induced 
electric field on $\Sigma$ in componenents along the 
normal and in the tangent plane. The transformation is given by $ x = X( {\xi^1 ,\xi^2 } ) + \xi^3 n_x ( {\xi^1 ,\xi^2 } )$, 
$y = Y( {\xi^1 ,\xi^2 } ) + \xi^3 n_y ( {\xi^1 ,\xi^2 } )$, and $z = Z( {\xi^1 ,\xi^2 } ) + \xi^3 n_z ( {\xi^1 ,\xi^2 } )$, 
where $n_x,\,n_y,\,n_z $ are, respectively, the $x - , y - $, $z - $components of the 
normal ${\bf n}$. Thus the induced electric field on $\Sigma $ is actually calculated in the neighborhood 
of $\xi^3 =0$. The vectors $({{\bf r}_{\xi^1 } ,{\bf r}_{\xi^2 } ,{\bf n}} )$ make a basis and a 
three-frame generated by the above nonlinear coordinate transformation. 
The basis $( {{\bf r}^{\xi^{1} } ,{\bf r}^{\xi^{2} } ,{\bf n}})$ that is dual to $( {{\bf r}_{\xi^1 } ,{\bf r}_{\xi^2 } ,{\bf n}})$ is given by

\begin{equation}
\label{eq24a}
{\bf r}^{\xi^{1} } = {\frac{ {\bf r}_{\xi^2} \times {\bf n}}{ {\bf n} \cdot ({\bf r}_{\xi^{1}} \times {\bf r}_{\xi^{1}})}},
\end{equation}

\begin{equation}
\label{eq24b}
{\bf r}^{\xi^{2} } = {\frac{{\bf n} \times {\bf r}_{\xi^1} }{ {\bf n} \cdot ({\bf r}_{\xi^{1}} \times {\bf r}_{\xi^{1}})}}.
\end{equation}

Then on $\Sigma$ the induced electric field along the normal $\bf n$ is given with the 
help of $\hat {M}$ as \cite{Putinar2007}

\begin{equation}
\label{eq25}
{\tilde E_n} ( {\xi^1 ,\xi^2 } ) = ( {\hat {M} + 1/2} )u = 
\sum\limits_k {\frac{n_k ( {\chi _k + 1/2} )}{\frac{1}{ {2\lambda }} - \chi _k }} u_k ( {\xi^1 ,\xi^2 } ).
\end{equation}

\noindent
From Eqs. (\ref{eq12}) and (\ref{eq17}) and from the expression of the gradient in the general curvilinear coordinates, \cite{Moon1988}
the rest of the induced electric field laying onto the tangent plane to 
$\Sigma $ has the following expression 

\begin{eqnarray}
\label{eq26}
 {\tilde {\bf E}_t} \left({\xi^1,\xi^2 }\right ) = - \nabla _t v \left( {\xi^1 ,\xi^2 
}\right) \quad \quad \quad \quad \quad \quad \quad \quad\quad \quad\quad \quad \quad  \\ 
 = - \sum\limits_k {\frac{n_k }{\frac{1}  
{2\lambda } - \chi _k }} [ {\frac{\partial v_k 
( {\xi^1 ,\xi^2 } )}{\partial \xi^1 }{\bf r}^{\xi^1 } + 
 \frac{\partial v_k ( {\xi^1 ,\xi^2 } 
)}{\partial \xi^2 }{\bf r}^{\xi^2 } }].  \nonumber
 \end{eqnarray}

\noindent
When the three-frame $\left( {{\bf r}_{\xi^1 } ,{\bf r}_{\xi^2 } ,{\bf n}} 
\right)$ is orthogonal, the induced field tangent to 
$\Sigma $ takes the form

\begin{eqnarray}
\label{eq26a}
 {\tilde {\bf E}_t} ( {\xi^1,\xi^2 } ) = \quad \quad \quad \quad \quad \quad \quad \quad \quad \quad \quad \quad \quad\quad \quad\quad \quad \quad\quad\\
 - \sum\limits_k {\frac{n_k }{\frac{1}  
{2\lambda } - \chi _k }} [ {\frac{1}{h_{\xi^1 } }\frac{\partial v_k 
( {\xi^1 ,\xi^2 } )}{\partial \xi^1 }{\bf t}_{\xi^1 } + 
\frac{1}{h_{\xi^2 } }\frac{\partial v_k ( {\xi^1 ,\xi^2 } 
)}{\partial \xi^2 }{\bf t}_{\xi^2 } }]. \nonumber
 \end{eqnarray}

\noindent
Equations (\ref{eq25}) and (\ref{eq26}) are the main results of this work. These 
equations provide an eigenmode decomposition of the near-field
and an intuitive and a direct relationship between the LSPRs and their local 
field enhancements. In the vicinity of $\Sigma$ the total near-field is the sum of 
the induced electric field ${\tilde {\bf E}}$ and the applied field ${\bf E}_0$: 
${\tilde {\bf E}}_{total} = {\tilde {\bf E}} + {\bf E}_0 =  {\tilde {\bf E}}_{t} + {\tilde E}_n \bf{n} + {\bf E}_0 $. 
The total electric field is a complex-valued quantity. Its modulus represents the strength of the total electric field and its 
phase is the phase shift between the applied and the total field. 

The near-field enhancement is 

\begin{equation}
\label{eq27a}
\frac {|{\tilde {\bf E}}_{total}|} {|{\bf E}_0 |} \cong 
\frac {|{\tilde {\bf E}}|} {|{\bf E}_0 |}
\end{equation}

\noindent
since ${|{\tilde {\bf E}}|}/{|{\bf E}_0 |}\gg 1$ at the plasmon resonance frequency. 
There are several consequences of these results. First, the equations (\ref{eq25}) 
and (\ref{eq26}) locate the hot spots of the near-field enhancement. The 
spatial maxima of the normal component of the near-field enhancement are 
provided by the maxima of the absolute value of $u_k $, whilst the maxima of 
the tangent component are localized in the regions of fast variations of 
$v_k $. Although $v_k$ is a smooth version of $u_k$ by (\ref{eq12}), the areas with fast variations of 
$v_k $ may come from the regions of rapid change of $u_k $.  Thus the simple inspection of the eigenfunctions indicates the regions with 
high near-field enhancement. Second, although there is a direct relation 
between the near- and far-field coupling to the electromagnetic radiation 
via ${w}_{k} = {n}_{k} \langle {\bf{x}} \cdot {\bf{N}}\vert {u}_{k}/{V}$, there are eigenmodes with large 
near-field enhancements but with small dipole moments, thus being almost dark in the far-field. 
Third, in the Drude metal case, Eqs. (\ref{eq25}) and (\ref{eq26} have a frequency-dependence form 
similar to Eq. (\ref{eq19}). Starting from the latter one can arrive at a fourth consequence related to the difference between the 
near- and far-field spectral properties. Recent works indicate a spectral 
shift between the maxima of the far- and near-field spectral response 
\cite{Bryant2008,Ross2009,Zuloaga2011}. From eqs. (\ref{eq19}) and (\ref{eq20}) the far-field spectral maximum of the 
$k^{th}$ LSPR is the maximum of the function

\begin{equation}
\label{eq28}
Im( {\alpha _{plas} } ) \propto \frac{w_k }{1 / 2 - \chi _k 
}\frac{\varepsilon _d }{\varepsilon _{eff\_k} } 
\frac{\tilde {\omega }_{pk}^2 \omega ^2}{( {\omega ^2 - \tilde {\omega }_{pk}^2 } )^2 + 
( {\omega \gamma } )^2},
\end{equation}

\noindent
whose maximum is at $\omega = \tilde {\omega }_{pk} $. On the other hand, if 
it is assumed that the $k^{th}$ LSPR is well resolved then the spectral 
maximum of the $k^{th}$ LSPR near-field is the maximum of

\begin{equation}
\label{eq29}
| {\bf E} | \propto \frac{n_k }{1 / 2 - \chi _k }\frac{\varepsilon _d 
}{\varepsilon _{eff\_k} }\tilde {\omega }_{pk}^2 \frac{1}{\sqrt {( {\omega ^2 - \tilde {\omega }_{pk}^2 } )^2 + ( {\omega \gamma })^2} }.
\end{equation}

\noindent
The maximum of (\ref{eq29}) is at $\omega = \sqrt {\tilde {\omega }_{pk}^2 - {\gamma^2/2}} $. The results provided by (\ref{eq28}) and (\ref{eq29}) 
explain in a general fashion the spectral shift between the far- and near-field 
without invoking a mechanical analogy of the plasmon 
resonance phenomenon \cite{Zuloaga2011}. Nanoparticles made of metals with larger damping constants $\gamma$ show larger and easier discernible shifts \cite{Chen2011}. 
In the next section I am going to analyze two numerical 
examples that show the utility of the BIE method in estimating the 
near-field enhancement. 

\begin{figure}
\includegraphics[width=3.5in]{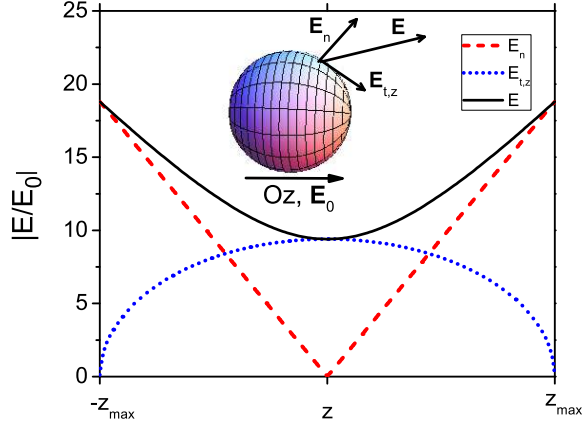}
\caption{\label{fig:1}
Spatial dependence of the near-field enhancement components for a metallic nanosphere in the $x-z$ plane at the resonance frequency. 
Normal component-red dashed line, tangent component-blue dotted line, total near-field 
enhancement-black solid line). All three quantities are axially symmetric about $z$-axis. The inset shows a nanosphere and the 
directions of the applied and induced fields. 
}
\end{figure}

\section{Numerical examples}
%\subsection{Numerical procedure}

In the numerical implementation of the method presented above I consider NP 
shapes with axial symmetry. The surface $\Sigma $ may be parameterized by 
equations like 
$\{x,y,z\} = \{g(z)\cos\phi,\; g(z)\sin\phi, \;z\}$ 
or $\{x,y,z\} = \{r(\theta)\sin\theta \cos\phi, \; r(\theta)\sin\theta \sin\phi, \; r(\theta) \cos\theta\}$, 
where $g(z)$ and $r(\theta)$ are smooth and arbitrary functions of 
$z$ and $\theta $, respectively. In the first case the independent 
parameters are $z$ and $\phi $, while in the second case the parameters 
are $\theta $ and $\phi $. These two parameterizations provide an orthogonal three-frame on $\Sigma$ and a smooth 
mapping to a standard sphere. Hence the basis functions in which the 
operators $\hat {M}$, $\hat {M}^{\dagger} $, and $\hat {S}$ are expressed are easily related
to spherical harmonics $Y_{lm}$ \cite{Sandu2010,Sandu2011}. The axial symmetry ensures some selection rules of the LSPRs. 
Thus for field polarization parallel to the symmetry axis the selection rules imply basis functions with $m=0$. In the same 
time, for a transverse polarization only the basis functions with $m=1$ give non-zero matrix elements. 
Numerical calculations are made with gold NPs immersed in water with a dielectric constant $ \varepsilon _{d}=1.7689$.  
The dielectric function of the gold NPs is adjusted such that the plasmon resonance wavelength of a gold nanosphere immersed in water is at 529 nm. The following Drude constants are used for gold: $\varepsilon_{m}=11.2$, $\hbar\omega_{p}=9$ eV, and $\hbar\gamma=100$ meV. 
The damping constant $\gamma$ incorporates the bulk damping and the damping due to surface collisions of electrons \cite{Sandu2011}.   
 
\begin{figure}[htp]
  \begin{center}
   \subfigure {\label{fig:2a} \includegraphics {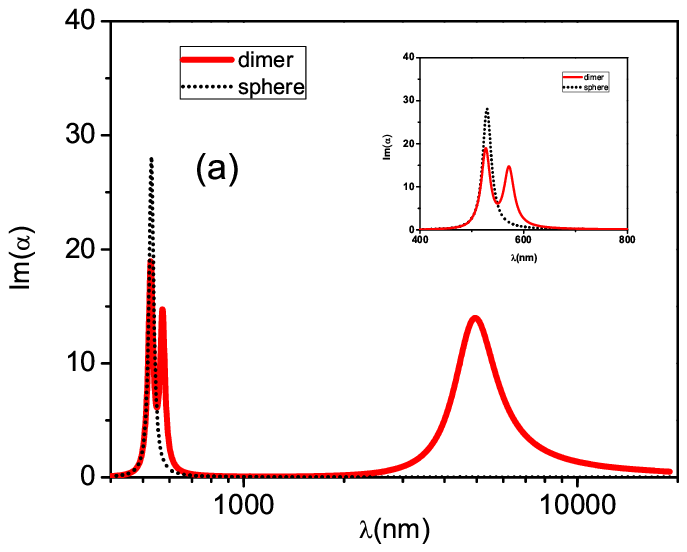}}  \\
    \subfigure {\label{fig:2b} \includegraphics {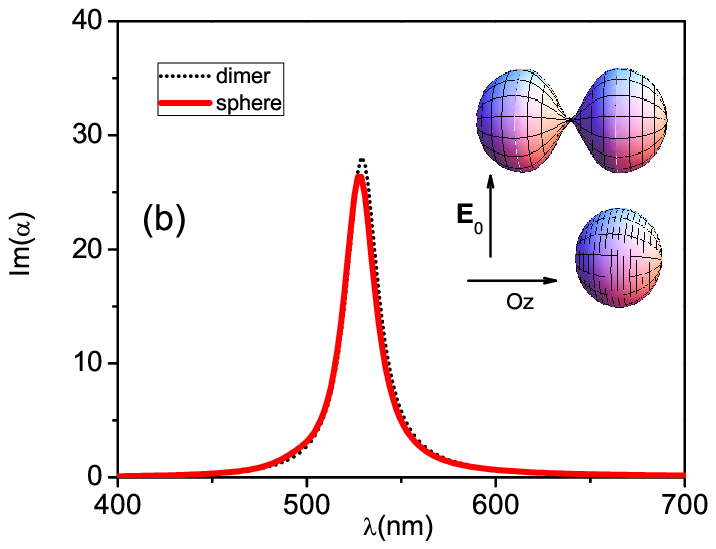}} 
  \end{center}
  \caption{Imaginary part of polarizability for a gold nanosphere (black dotted line) and 
for a dimer of nearly spherical NPs (red full line) in (a) visible and infrared for an applied field parallel to $z$-axis and in (b) visible 
for the polarization perpendicular to $z$-axis. The scaling of the polarizability by the NP volume makes it a dimensionless quantity. The inset 
of (a) shows the imaginary part of polarizability only in visible and the inset of (b) shows the shapes of the nanosphere and the dimer.}
  \label{fig:2}
\end{figure}

\subsection{Metallic nanosphere and spherical dimer in parallel field}
The dimers exhibit large near-field enhancement \cite{Atay2004,Rechberger2003,Su2003} and, in particular, the nearly touching dimers reveal also a resonance in infrared part of the spectrum \cite{Romero2006,Sandu2011}. In this subsection I examine and compare the plasmon resonance properties of spherical NPs and of nearly touching dimers made of almost spherical particles. Although a nearly touching dimer proves to be difficult to fabricate, it may model a system closely related to those that have large SERS enhancement like the nanostars deposited onto a smooth gold surface \cite{Lorenzo2009}. A sphere 
presenting a tip close to a gold film and showing a large near-field enhancement \cite{Puebla2010} may have a correspondent in a nearly touching dimer due to the image charge that appears like another particle supporting LSPRs \cite{Vernon2010}.   

The sphere surface can be described by the equation $\{x,y,z\} = \{g(z)\cos\phi,\; g(z)\sin\phi, \;z\}$, where 
$g(z) = z_{max}\sqrt{1-z^2/z^2_{max}} $ and $z_{max}$ is the radius of the sphere. The shape of a dimer made of nearly spherical 
particles connected by a tight junction is taken from a more general equation for clusters of $n$ touching particles having the same parametrization
$\{x,y,z\} = \{g(z)\cos\phi,\; g(z)\sin\phi, \;z\}$, where \cite{Sandu2011}

\begin{eqnarray*}\label{dimer_shape}
g(z) = 
A( 1 + S(zS(z) - (n-1) a) ) \times \\
\frac{\sqrt{1 - ((z - S(z)(n-1)a)/a)^2}}
{1 + (1 + b(z - S(z)(n-1)a)^2)^2} 
[ 1 - Fl( \frac{S(z)z + a}{na})]\times\\
\left[ h + 2(A-h) ( 1 - \frac 1
{1 + (1 - (H(z)/a)^2)^2})\right]
\end{eqnarray*}
with $H(z)=Mod((-1)^{Fl(z/a)+n-1}z, a)$;
$S(z)$ is signum function and equals -1, 0 or 1 if z is negative, zero or positive;  $Fl(z)$ is the greatest integer less than $z$; and
$Mod(x,y)$ is the remainder of the division of $x$ by $y$.
Parameters $A$ and $a$ define the radius of the maximum cross-section and the half-length of any particle in the cluster, respectively.  
Thus the ratio $a/A$ is the aspect ratio of a particle in the cluster. Parameter $b$ determines the curvature of the end caps such that 
for spherical end caps $b=0$- while $h$ gives the coss-section size of the connecting gap. For a nearly spherical 
dimer $n=2$, $a=A=z_{max}$, $b=0$, and $h=0.025z_{max}$.

The first example examined is the nanosphere, whose field enhancement is a textbook calculation \cite{Maier2007}. 
The near-field enhancement of a nanosphere  in the $x-z$ plane at resonance frequency is presented in 
Fig. \ref{fig:1}. In the $x-z$ plane the $x-$coordinate is determined by the equation $x=h(z)$. The field polarization 
is parallel to $z-$axis, therefore the induced field is also symmetric about $z-$axis.  A comparison of far-field spectra 
for the nanosphere and for the dimer is given in Fig. \ref{fig:2}. The bright eigenmode is the dipole mode corresponding to 
the second largest eigenvalue 
$\chi_2=1/6$ with $w_2=1$ (see also Table \ref{paraex} ). 
Its corresponding 
eigenfunctions $u_2(z),v_2(z) \propto z$. Therefore in Fig. \ref{fig:1}, the normal component of 
$|\bf{E}/\bf{E}_0|$ is linear in $z$, while the tanget component acquires the $z-$dependence of $1/h_z(z)$, where $h_z(z)$ is 
the Lam\'{e} coefficient for the independent parameter $z$. Along z-axis the maximum near-field enhancement is about 19 
occuring at the north and south poles of the nanosphere.

\begin{table}
\caption{The most representative eigenvalues, their plasmon resonance wavelengths, and their weights $w_k$ and $n_k$ for a sphere and a dimer made of nearly spherical particles connected by 
a tight junction. The field is parallel ($\bf{E}_0 || Oz$ and $m=0$) or perpenidcular($\bf{E}_0 \perp Oz$ and $m=1$)  to symmetry axis.}
\label{paraex}
\begin{ruledtabular}
\begin{tabular}{lc|cccc|c|cccc}
&  &sphere &     &   &     &  &dimer &    &   &     \\
\hline
   &$k$ & $\chi _{k}$ & $\lambda_k $(nm) &   $w_{k}$ & $n_{k}$ & $k$ & $\chi _{k}$ &$\lambda_k (nm)$ &   $w_{k}$  & $n_{k} $  \\
\hline
& &  $m=0$  &    &     &     &     &   $m=0$    &     &     &        \\
\hline
& 1 & 0.5 &$\infty $  &    0.0 &0.0 & 1 & 0.5&$\infty $  &    0.0 &0.0  \\
&      &     &      &      &  & 2& 0.498571 &  4867 &    0.0071&  -0.025\\
\hline
&   &      &  &      &      & 3 &  0.272&  571.6  &    0.287 & -1.97 \\
\hline
&  &     &    &      &      & 4 &  0.201& 540 &    0.0 &  0.0   \\
\hline
& 2  & 0.167&  529   &  1& -3.34  & 5&   0.159& 526.9 &    0.658&  -3.66 \\
&    &     &    &      &      & 6 & 0.122& 517.8 &    0.0& 0.0   \\
\hline
& 3   &  0.1& 513  &    0.0&  0.0  & 7 & 0.112& 515.5 &    0.027&  -0.79 \\
&    &      &    &       &     & 8 & 0.067&  506.6 &    0.0& 0.0   \\
\hline
& &  $m=1$  &   &  &           &    &      $m=1$    &      &    &           \\
\hline
&1 &  0.167& 529 &   1& -2.36  &1 &  0.214 & 544.5 &    0.0& 0.0  \\
& &     &    &      &      & 2 & 0.162& 527.8&    0.96&  -3.09 \\
\hline
&  &    &    &       &     &3 &  -0.014& 494.5 &    0.022& 0.557 \\
\hline
&  &    &    &       &     & 4 & -0.086& 486.3 &    0.01& -0.408 \\
\end{tabular}
\end{ruledtabular}
\end{table}

\begin{figure*}[!tbp] %[htp]
  \begin{center}
   \subfigure {\label{fig:3a} \includegraphics [width=3.in]{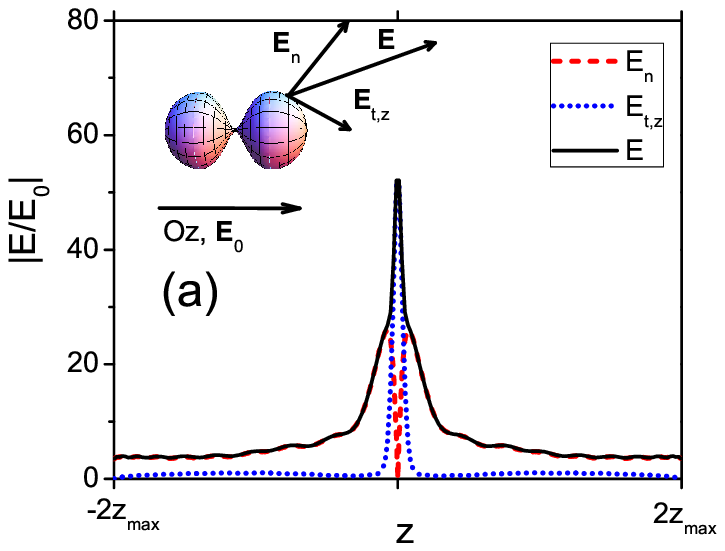}} 
    \subfigure {\label{fig:3b} \includegraphics [width=3.in]{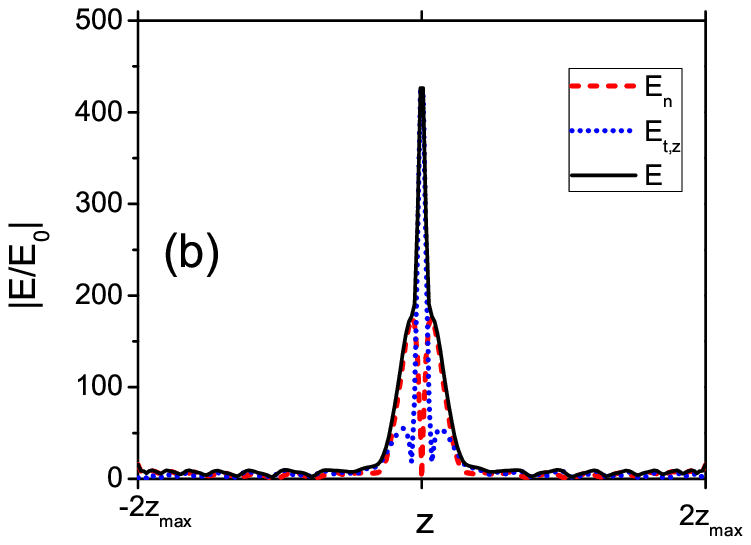}}  \\
     \subfigure {\label{fig:3c} \includegraphics [width=3.in]{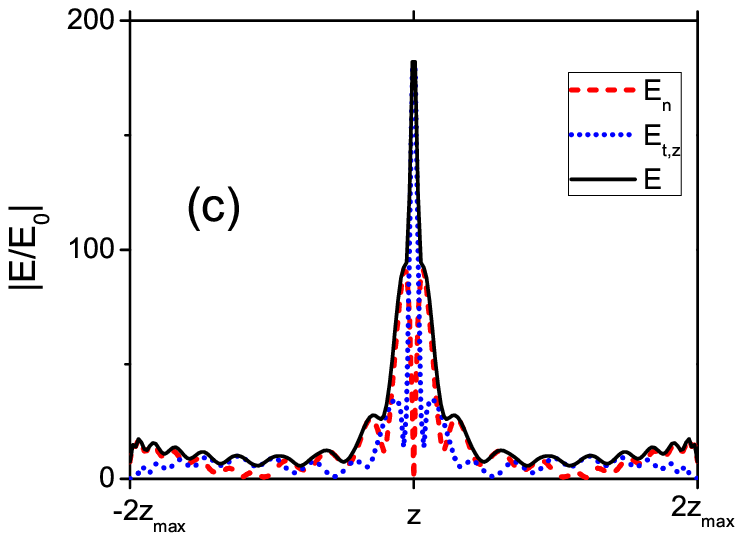}} 
    \subfigure {\label{fig:3d} \includegraphics [width=3.in]{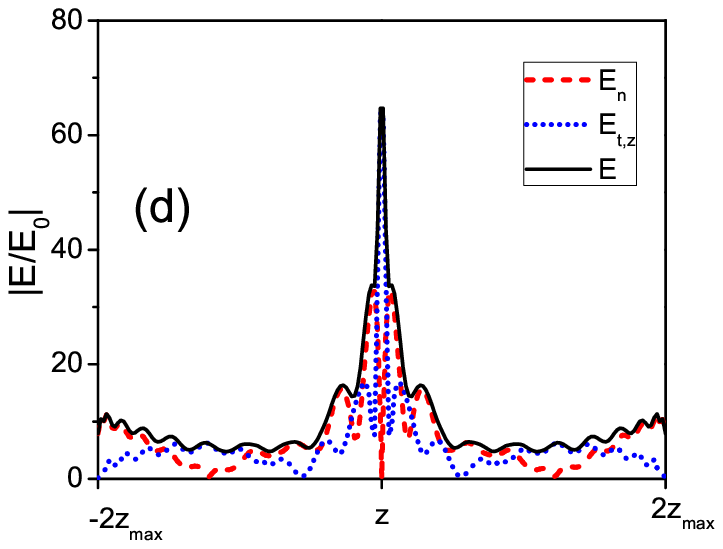}} 
  \end{center}
  \caption{The near-field enhancement in the $x-z$ plane at four plasmon resonance wavelengths given by: (a) the 
second ($\lambda=4867$ nm), (b) the third ($\lambda=571.6 $ nm), (c) the fifth ($\lambda=526.9$ nm), and 
(d) the seventh ($\lambda=515.5$ nm) dimer eigenmode from Table \ref{paraex}.   
The normal component of the enhancement is plotted by red dashed line, the tangent component by 
blue dotted line, and the total enhancement by black full line. The polarization of the field is parallel 
to the symmetry axis of the dimer, hence all three fields have axial symmetry. The inset of (a) shows the components of the 
near-field induced at the dimer surface.}
  \label{fig:3}
\end{figure*}

In Table \ref{paraex}  there are presented the most 
representative eigenvalues of both the nanosphere and the dimer, while  the far-field spectral behavior is plotted in 
Fig. \ref{fig:2}. The comparative analysis of the far-field spectrum has revealed that, with respect to a single 
sphere, the dimer has two more LSPRs in addition to that corresponding to the nanosphere alone: one in visible at longer 
wavelengths and the other more displaced into mid-infrared \cite{Sandu2011}. The eigenfunctions of the sphere and of the dimer are 
plotted in the Supplementary Information. The representative eigenmodes of the dimer are either hybrid modes of 
the nanosphere eigenmodes or proper eigenmodes of the dimer. Thus the first eigenmode of the nanosphere ($k=1$) has two 
corresponding hybrid eigenmodes in the dimer ($k=1$ and $k=2$): one is a symmetric combination of the nanosphere 
eigenmodes and the other is an anti-symmetric combination. Only the anti-symmetric mode is a bright eigenmode. In general, 
if a sphere eigenmode is symmetric under the mirror symmetry $z \leftrightarrow -z$, an anti-symmetric hybridization 
leads to a bright eigenmode of the dimer. Conversely, if a sphere eigenmode is anti-symmetric under the mirror 
reflection $z \leftrightarrow -z$, a symmetric hybrid can couple with the light. Consequently, the fifth and 
the sixth dimer eigenmodes are the hybrids of the second sphere eigenmode and the third sphere eigenmode creates the 
seventh and the eighth dimer eigenmodes. 

The third and the fourth dimer eigenmodes are characteristic to dimer itself. From the Supplementary Information 
one can see that the third eigenmode exhibits just a large dipole at the junction and therefore is bright. 
This mode has been noticed  in clusters of touching nanoparticles with $A/a \le 1$ \cite{Sandu2011}. On 
the other hand, the fourth eigenmode has a large charge accumulation at the junction but is even with 
respect to the mirror symmetry $z \leftrightarrow -z$, being therefore dark. In the terminology of 
Refs. \onlinecite{Klimov2007a} and \onlinecite{Klimov2007b} the hybrid modes are "atomic" modes, while the modes 
like the third and the forth eigenmode are called "molecular" modes. On the whole, all bright eigenmodes manifest large 
charge accumulations and fast changes of electric potential at the junction. According to Eqs. (\ref{eq25}) 
and (\ref{eq26}) at the resonance wavelengths there are huge near-field enhancements around the junction 
of the dimer as depicted in Fig. \ref{fig:3}. The enhancement is mostly provided by the normal 
component of the electric field, excepting the middle of the dimer, where the normal field vanishes 
and the tangent component contributes to the enhancement. 
The weights $w_2$ and $n_2$ of the second eigenmode are rather modest. Still, at $\lambda=$ 4867 nm 
the mode has a top near-field enhancement of 55, as these weights are magnified by 
the factor $1/(0.5 - \chi_2)$, which is huge for the corresponding eigenvalue.    
The rest of the eigenmodes have the the near-field enhancement maxima of 450 for the third eigenmode, 
of 180 for the fifth, and of 65 for the seventh   dimer eigenmode at their corresponding resonance 
wavelengths. All these field enhancements are much larger than the enhancemnt of a single nanosphere.

\begin{figure*}[!tbp] %[htp]
  \begin{center}
   \subfigure {\label{fig:4a} \includegraphics [width=3.in]{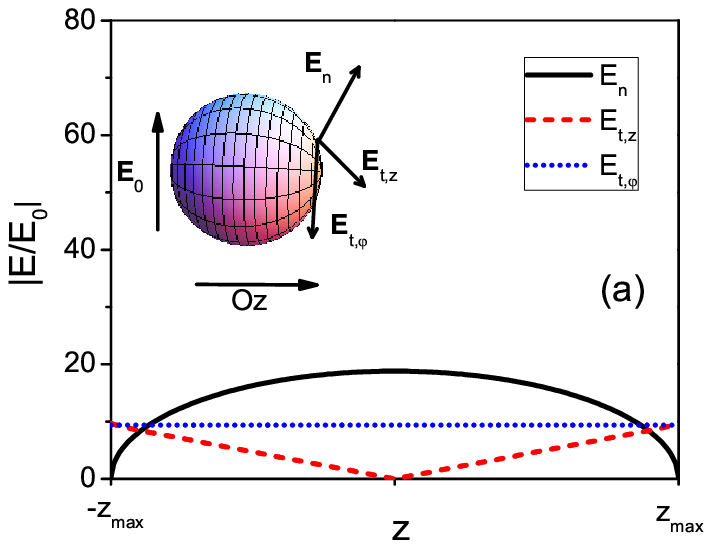}} 
    \subfigure {\label{fig:4b} \includegraphics [width=3.in]{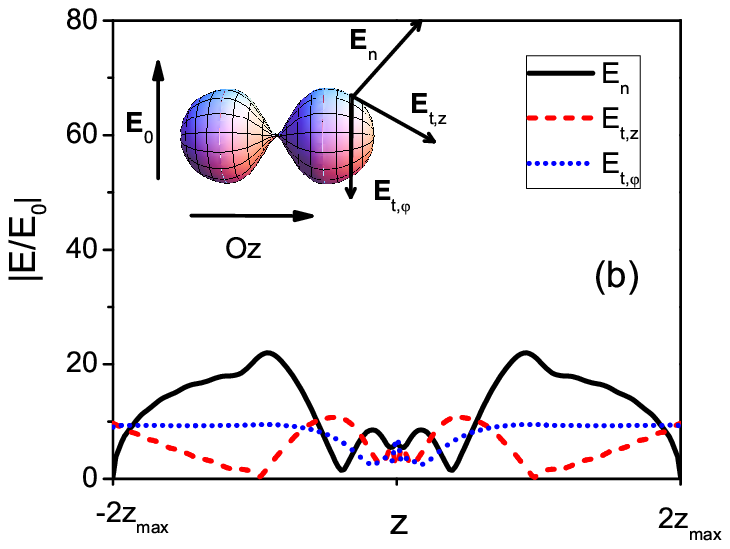}}  \\
     \subfigure {\label{fig:4c} \includegraphics [width=3.in]{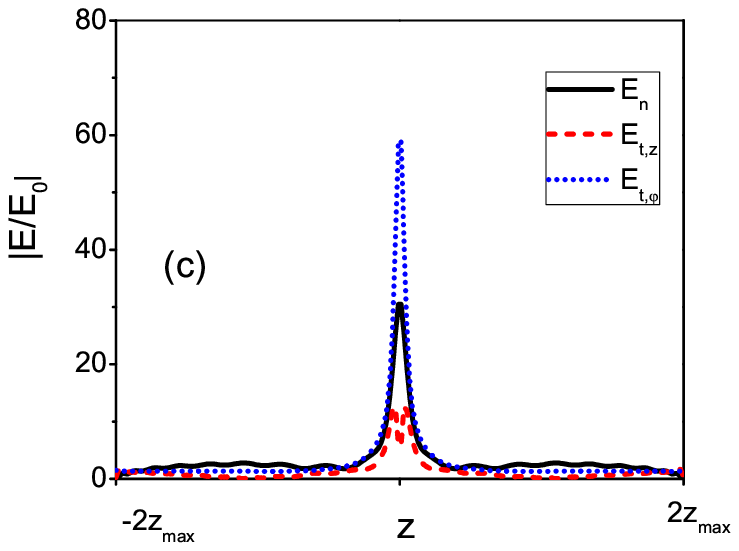}} 
    \subfigure {\label{fig:4d} \includegraphics [width=3.in]{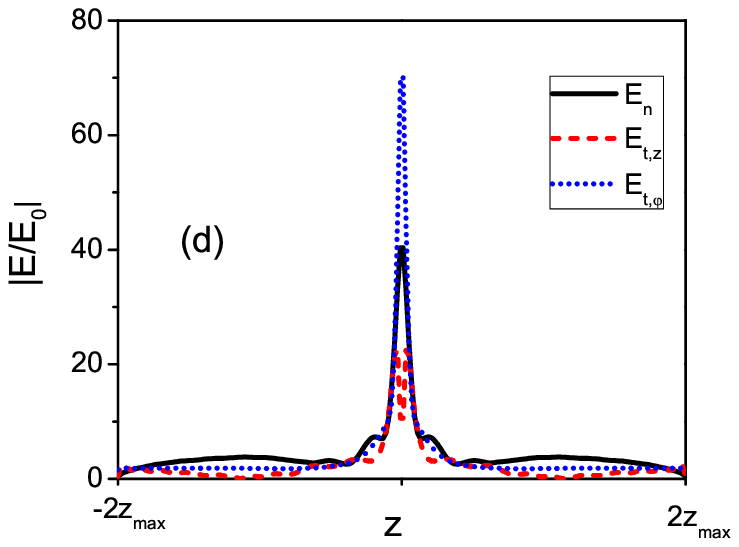}} 
  \end{center}
  \caption{The $z-$dependence of the near-field enhancement in transverse field for (a) the 
nanosphere at $\lambda=529$ nm, (b) the dimer at $\lambda=527.8$ nm, (c) the dimer at $\lambda=494.5$ nm, and 
(d) also the dimer at $\lambda=486.3$ nm.   
The normal component of the enhancement ${\bf E}_n$ is plotted by black solid line, the first tangent component ${\bf E}_{t,z}$ by 
red dashed line, and the second tangent component ${\bf E}_{t,\phi}$ by blue dotted line. The field components ${\bf E}_n$ and ${\bf E}_{t,z}$ have an additional multiplicative factor $cos(\phi)$ while the component ${\bf E}_{t,\phi}$ has the factor  $sin(\phi)$. The insets of (a) and (b) show schematically all three components of the near-field for the sphere and for the dimer, respectively.}
  \label{fig:4}
\end{figure*}

\subsection{Metallic nanosphere and spherical dimer in transverse polarization field}
In contrast to the behavior in parallel field the LSPRs present different characteristics in transverse (field) polarization. 
The far-field spectrum presented in Fig. \ref{fig:2}b shows similar spectrum for both types of NPs. The dimer resonance gets slightly 
smaller and slightly blue-shifted with respect to the resonance of a sphere \cite{Sandu2011}. This can be also seen 
from Table \ref{paraex}. Also the eigenmodes of the dimer are either hybrids of the sphere modes or proper dimer modes which are localized 
at the junction. These eigenmodes can be inspected comparatively to those of the sphere in the Supplementary Information. Conversely to the 
hybridization along the symmetry axis, in the transverse field a symmetric (even) combination of the sphere dipole modes leads to a bright 
dimer hybrid mode. In addition to that there is no charge accumulation at the junction with no additional near-field enhancement with respect 
to the sphere. The field enhancement for sphere and dimer are plotted in Fig. \ref{fig:4}, where only the $z-$dependence is represented. 
Due to the surface parameterization, the near-field enhancement of the sphere shown in Fig. \ref{fig:4}a does not appear to look similar to the 
field enhncement depicted in Fig. \ref{fig:1} even though they represent the same field. In Fig. \ref{fig:4} the field 
components ${\bf E}_n$, ${\bf E}_{t,z}$, and ${\bf E}_{t,\phi}$ have also a $\phi-$dependence as follows. ${\bf E}_n$ and ${\bf E}_{t,z}$ acquire 
the factor $cos(\phi)$ while ${\bf E}_{t,\phi}$ has an additional $sin(\phi)$ as a factor. At the extremities of the dimer and of the sphere the 
near-field enhancements of the hybrid dipoles and of the sphere dipole, respectively are equal, while they differ significantly at the junction region 
(Figs. \ref{fig:4}a and \ref{fig:4}b). The 
large near-field enhancement comes from the proper dimer eigenmodes that are basically dark in the far-field (Figs. \ref{fig:4}c and \ref{fig:4}d 
and Table \ref{paraex}) but have enhancements of 65 and 75, repectively at the junction (Figs. \ref{fig:4}c and \ref{fig:4}d and Table \ref{paraex}). 
Thus these two proper eigenmodes act as merely evanescent modes.     

\begin{table}
\caption{The most representative eigenmodes, their plasmon resonance wavelengths, and their weights $w_k$ and $n_k$ for a prolate spheroid and a nanorod with an aspect ratio of $5:1$. The field is parallel to the symmetry axis ($\bf{E}_0 || Oz$ and $m=0$).}
\label{pro_cyl}
\begin{ruledtabular}
\begin{tabular}{lc|cccc|c|cccc}
&  &spheroid &     &   &     &  &nanorod &    &   &     \\
\hline
   &$k$ & $\chi _{k}$ & $\lambda_k $(nm) &   $w_{k}$ & $n_{k}$ & $k$ & $\chi _{k}$ &$\lambda_k (nm)$ &   $w_{k}$  & $n_{k} $  \\
\hline
& &  $m=0$  &    &     &     &     &   $m=0$    &     &     &        \\
\hline
& 1 & 0.5 &$\infty $  &    0.0 &0.0 & 1 & 0.5&$\infty $  &    0.0 &0.0  \\
\hline
&  2    &  0.44418   & 884.8     &   1   &1.08  & 2& 0.4481&  909.5 &    0.9&  1.17\\
\hline
&   &      &  &      &      & 3 &  0.280&    &    0.059& 0.616\\
\hline
&  &     &    &      &      & 4 &  0.174&    &    0.019&  -0.425\\
\hline
&   &   &      &   &     & 5&   0.121&     & 0.014&  -0.394\\
&    &     &    &      &      & 6 & 0.098&    &    0.009& 0.327\\
\end{tabular}
\end{ruledtabular}
\end{table}

\subsection{Nanorod versus prolate spheroid in parallel field}
\begin{figure*}[!tbp] %[htp]
  \begin{center}
   \subfigure {\label{fig:5a} \includegraphics [width=3.in]{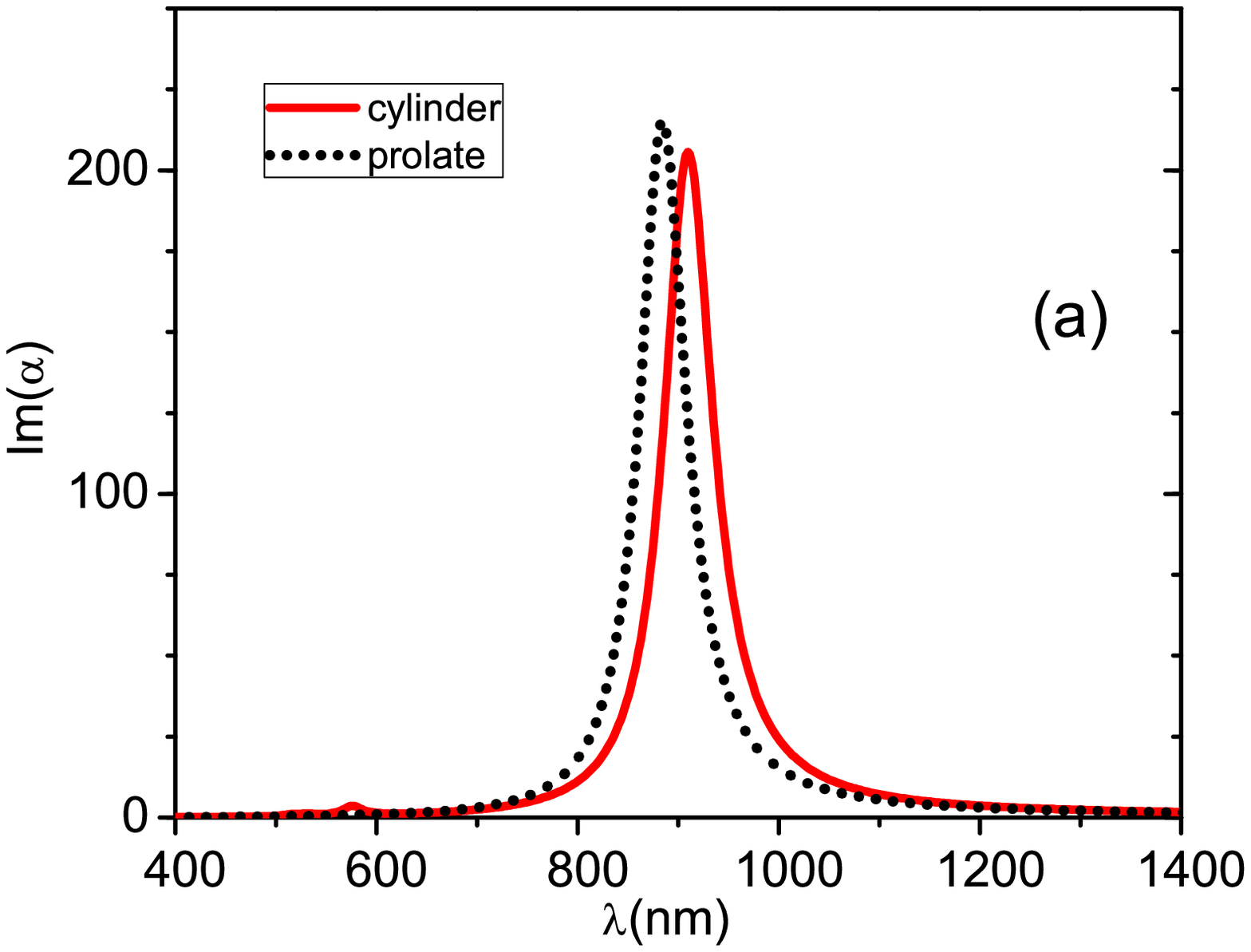}} 
    \subfigure {\label{fig:5b} \includegraphics [width=3.in]{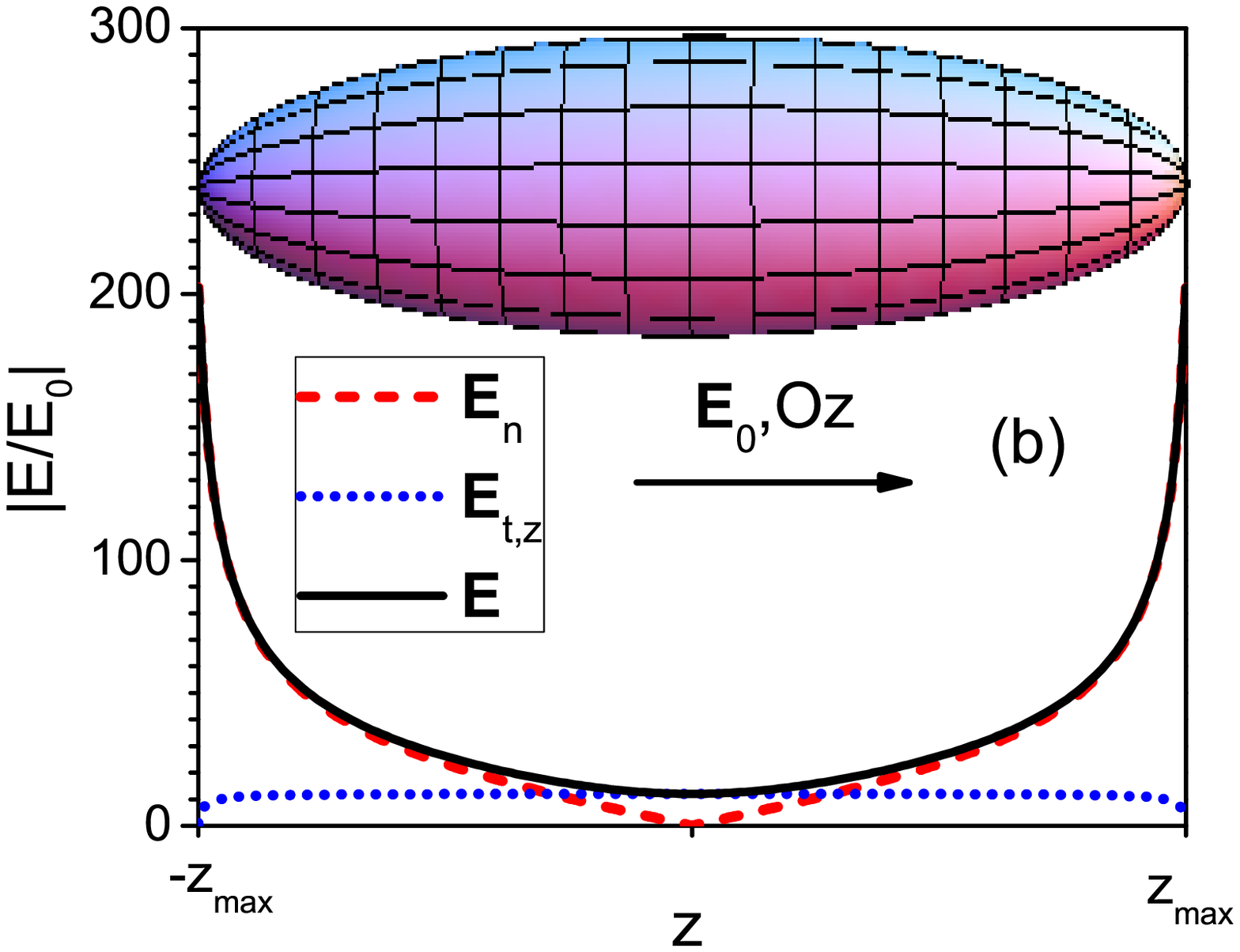}}  
     \subfigure {\label{fig:5c} \includegraphics [width=3.in]{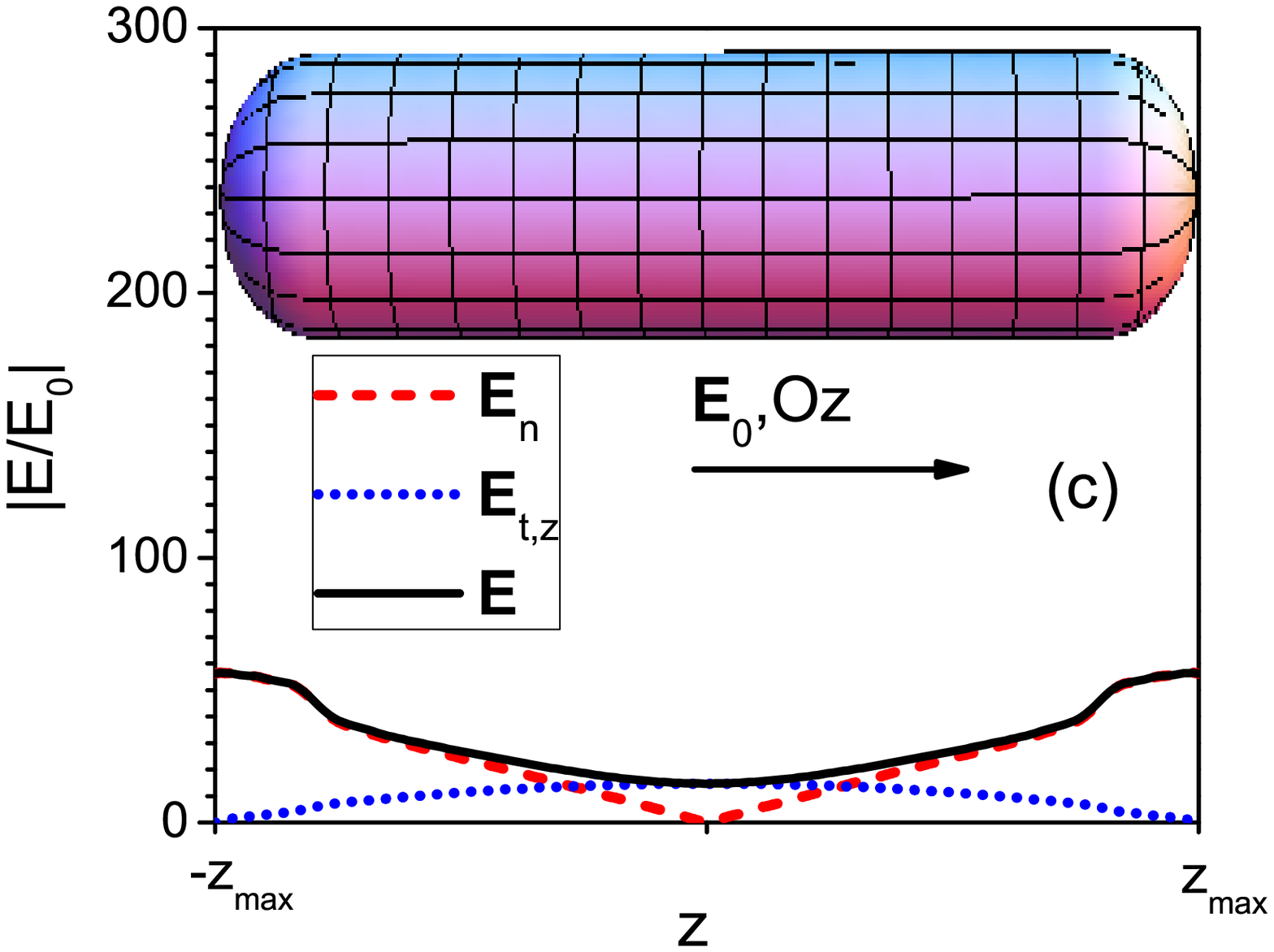}} 
  \end{center}
  \caption{(a) The far-field spectral behavior of a prolate spheroid (black dotted line) and of a nanorod (red solid line) of the same aspect ratio $5:1$ in parallel field; the near-field enhancement of (b) the prolate spheroid at $\lambda=884.8$ nm, and (c) of the nanorod at $\lambda=909.5$ nm in the $x-z$ plane.    
The normal component of the enhancement ${\bf E}_n$ is plotted by red dashed line, the tangent component ${\bf E}_{t,z}$ by blue dotted line, and the total enhancement ${\bf E}$ by black solid line. Due to the parallel field polarization all three fields have axial symmetry.}
  \label{fig:5}
\end{figure*}

For more than a decade a large number of wet chemistry methods have been developed for synthesis of metal NPs in a wide range of shapes and 
sizes\cite{Odom2008}. Of great interest are metallic nanorods due to the flexibility of controlling their aspect ratio, hence controlling their 
spectral response over the entire range of visible spectrum as well as in the near infrared. In general, larger aspect ratio implies larger eigenvalues and larger plasmon resonance wavelengths \cite{Sandu2011}. 
Commonly, metallic nanorods have been modeled as prolate spheroids due to their spectral response, which can be modeled analytically and, 
as it turns out, is quite close to the spectral response of a nanorod. Here I consider cylindrical nanorods capped with half-spheres and prolate spheroids with the same aspect ratio like those of the nanorods. The same type of parameterization $\{x,y,z\} = \{g(z)\cos\phi,\; g(z)\sin\phi, \;z\}$ is used for spheroids and nanorods. The spheroids are defined by a function $g(z)$ of the form $g(z) = \sqrt{1-z^2/z^2_{max}}$, where the aspect ratio is $z_{max}:1$. In a similar manner one can also define the nanorod shape with the same aspect ratio. In numerical calculations a $5:1$ aspect ratio is used. 
Table \ref{pro_cyl} presents the representative eigenmodes for the both types of NPs in parallel field polarization. Their far-field spectrum is 
presented in Fig. \ref{fig:5}a. The second largest eigenvalue $\chi_2$  gives the main plasmon response in both cases and the 
difference appears only at the third digit. Furthermore, its weights $w_2$ and $n_2$ are also quite similar. 
However, Figs. \ref{fig:5}b and \ref{fig:5}c show that the near-field enhancement at the ends of the prolate spheroid is almost four times larger than the near-field at the ends of the nanorod (202 versus 56). The large difference in the near-field enhancement comes from the corresponding eigenfunction $u_2(z)$, which gives the spatial dependence of the field enhancement. The eigenfunction $u_2$ is affected at its maximum by the curvature at the ends, where the spheroid has a different curvature from that of the nanorod. Hence this result suggests that increasing the near-field enhancement requires local changes of shape in the region where the eigenfunction $u_k(z)$ reaches its absolute value maximum .

%In table \ref{paraex} 

\section{Conclusions}

In this work I present a powerful and intuitive technique that relates directly the near-field enhancement factor to the eigenvalues and eigenvectors 
associated with the BIE method. Similarly to the far-field, the near-field is expressed as a sum over the eigenmodes of the BIE operator. 
This property offers 
a general explanation of the spectral shift between the far- and near-field maxima. The current method allows fast identification of near-field hot spots 
just by inspecting the eigenfunctions $u_k $ of the BIE operator and $v_k $ of its adjoint. The normal component of the near-field enhancement peaks in the 
regions where the absolute value of $u_k $ attains its maxima. Moreover the maxima of the tangent component of the near-field enhancement are found in 
the regions with fast variations of $v_k$. Although $v_k$ is smoothed-down through Eq. (\ref{eq12}), one can detect fast variations of $v_k$ by looking 
for fast variations of $u_k$. In addition to that, Eq. \ref{eq12} provides a physical meaning to $v_k$ as the electric potential generated by the charge 
distribution $u_k$. 

The procedure is applied to several types of NPs which exhibit large near-field enhancement. The analysis of these examples shows the strength of the 
current method. The first example is the dimer of nearly touching spheres. The dimer eigenmodes are either hybrids of the spherical eigenmodes or proper 
dimer eigenmodes. The hybrid eigenmodes exhibit charge build-up at the junction only in the parallel field polarization.  The proper dimer eigenmodes, 
on the other hand, show strongly localized behavior at the junction in both polarizations. Thus the large near-field enhancement occurs via the huge 
charge accumulation and the fast potential change at the dimer junction. 
The second example treats the comparative near-field behavior of a nanorod and of a prolate spheroid of the same aspect ratio. These types of 
nanoparticles have similar far-field spectra but the near-field enhancement of the spheroid is almost four times higher at the field-oriented 
extremities of the nanoparticle. The latter clarifies that, in order to improve the near-field factor, one must bring targeted local corrections 
to the geometry by focusing on the regions where the absolute value of $u_k$ reaches its maximum. The current methodology can be easily extended to 
more complex systems like assemblies of NPs.

\begin{acknowledgments}
This work has been supported by the Sectorial Operational Programme Human Resources Development, 
financed from the European Social Fund and by the Romanian Government under the contract number POSDRU/89/1.5/S/63700. 
\end{acknowledgments}

%\bibliography{NFE_Titus_Sandu}


\end{document}